# Vortex stabilization in magnetic trilayer dots

P. Szary[a], O. Petracic and H. Zabel

Institut für Experimentalphysik, Ruhr-Universität Bochum, D-44780 Bochum, Germany

The magnetization reversal and spin structure in circular Co/insulator/$Ni_{80}Fe_{20}$ trilayer dots has been investigated numerically. The effect of dipolar coupling between a soft ferromagnetic Permalloy (Py=$Ni_{80}Fe_{20}$) layer and a hard ferromagnetic Cobalt layer inside one stack is studied. We find either a stabilization or even a triggering of the vortex state in the Py layer due to the magnetic stray field of the Co layer, while the Co magnetization remains in a single-domain state. Furthermore, for thin Py layers a 360°-domain wall is observed. We construct a phase diagram, where regions of vortex stabilization, triggering, and occurrence of a 360° domain wall are marked.

[a] corresponding author: philipp.szary@ruhr-uni-bochum.de

Recently, magnetic tunnel junctions (MTJs) have gained increasing importance in magnetic data storage applications [1-5]. In particular, tunnel magneto-resistance (TMR) elements play an important role, e.g. in magneto-resistive read heads or magnetic random access memories [6]. A TMR element consists of two ferromagnetic (FM) layers separated by an insulating spacer. One geometry of the ferromagnetic layer has attracted considerable attention: disk-shaped soft FM nanostructures. Those can exhibit a so-called vortex state, which is characterized by an almost completely flux-closed curled magnetization state within the plane surrounding a small singularity at the center of the vortex where the magnetization is tilted out-of-the plane, viz. the vortex core [7, 8]. The flux-closed configuration leads to negligible stray-fields and thus reduces problems with inter-element interactions in close packed arrays e.g. in data storage systems [9]. Another interesting property is that each vortex-state carries two bits of information simultaneously, i.e. the chirality (sense of rotation) and the polarity of the vortex-core pointing either up- or downward [7, 10-14]. Therefore, possible technical realizations are vividly discussed [15, 12], e.g. the vortex random access memory (VRAM) [12].

Although the static [8, 16-22] and dynamic [23-27] properties of single-layer FM disk-shaped nanostructures have been intensely investigated, the effect of inter-layer interactions onto the vortex state is scarcely studied [28, 29]. To address this issue we investigate cylindrical Cobalt/insulator/Permalloy (Co/ IS/ Py) nanostructures (dots) with various thicknesses of the Co- and the Py- layer using micromagnetic simulations. We find surprisingly a *stabilization* and even a *triggering* of the vortex state due to the dipolar coupling to the hard ferromagnetic (HFM) Co layer. For thin Py-thicknesses a 360°-domain wall (360°-DW) is observed. Furthermore, we can construct a phase diagram for different reversal mechanisms depending on the Co- and Py- layer thicknesses.

The investigated system is schematically depicted in the inset of Fig. 2. It is a disk-shaped trilayer dot which consists of a HFM Co-layer with thickness $t_{Co}$ and a soft ferromagnetic (SFM) Py-layer of thickness $t_{Py}$ separated by vacuum representing an insulating barrier of constant thickness 3 nm. The diameter is $d = 300$ nm. The thicknesses of the Co and Py layers are varied between 0 and 30 nm. We use the OOMMF micromagnetic simulation code [30] with the following parameters. The simulation cell size has been kept constant throughout as 3 nm × 3 nm × 3 nm. The anisotropy constants used are $K_{Co} = 0.5 \cdot 10^6$ J/m$^3$ for Co [31] and zero for Py. The exchange constants are $A_{Co} = 10 \times 10^{-12}$ J/m [31] and $A_{Py} = 9 \times 10^{-12}$ J/m [32, 33] and the saturation magnetization $M_{s,Co} = 1.44 \times 10^6$ A/m [31] and $M_{s,Py} = 8 \times 10^5$ A/m [34]. The damping constant is taken to be $\alpha = 0.5$. One should note that the Co- and the Py-layer are coupled only via the dipolar (demagnetization) energy. Hysteresis loops are recorded within the field range -3 kOe $\leq H \leq$ 3 kOe. In this range only the Py-layer shows a magnetization reversal, while the Co-layer remains saturated along the +$H$-direction. The switching of the Co-layer occurs at ±5 kOe.

Fig. 1 shows $M(H)$ hysteresis loops for three different cases: Fig. 1 (a) and (b) display the first case of *stabilization* of the vortex state in the Py-layer. Panel (a) shows the magnetization hysteresis loop $M(H)$ of a Py-dot with $t_{Py} = 30$ nm *without* Co. It shows a hysteresis curve characteristic for a vortex state [8, 35], which is also evidenced by the corresponding spin structure at $H = -0.3$ kOe [lower image in panel (a)] being the nucleation field of the vortex. The upper image at $H = -0.15$ kOe shows the spin structure directly before the system develops a vortex state. One clearly observes a single domain state with an onion-like bending of the spin structure due to the circular shape of the dot. This is to be compared to the case shown in panel (b), where also $t_{Py} = 30$ nm but with a Co-layer of $t_{Co} = 3$ nm. The hysteresis loop shows a more pronounced vortex state as seen from the collapsed central part of the $M(H)$-curve. The finding of vortex stabilization is surprising, because studies on the vortex state in FM dots usually show a *destabilization* of the vortex, e.g., due to anisotropy [36, 37], coupling [28, 29] or exchange biasing [38]. Few cases exist where a stabilization of the vortex state has been observed [39]. The insets show the corresponding spin structures again at the field value of vortex nucleation, i.e. $H = 0.15$ kOe [lower image] and at $H = 0.30$ kOe [upper image]. Here the latter clearly differs from a single domain state present in the case without Co. Such kind of spin configuration has been found before in other studies on Py monolayer dots both numerically [19, 20, 40] and experimentally [41]. The present state can be described as a so called "W-state" which is one of the various metastable intermediate buckling states between the vortex and onion state. The minor hysteresis loop of the Py layer appears vertically shifted due to the virtually constant magnetization contribution of the Co layer. Moreover, the dipolar interaction to the Co layer induces also a bias shift to positive field values.

In the second case, displayed in Fig. 1 (c) and (d), even a *triggering* of the vortex state can be observed, i.e., where a vortex state in the Py layer can be found only when the Co layer is present. Panel (c) shows the reference curve $M(H)$ of a Py-dot with $t_{Py} = 21$ nm *without* Co. Here the magnetization reversal occurs by direct switching without any vortex, which is evidenced by the spin structure images at $H = -0.3$ kOe [upper image] and $H = -0.45$ kOe [lower image]. The hysteresis curve shows a square-like loop. In comparison, the system together with a Co-layer of thickness $t_{Co} = 3$ nm displays a vortex as depicted in the insets of panel (d) for $H = 0.15$ kOe [lower image]. Again in the image at $H = 0.30$ kOe a W-state [upper image] can be recognized, which in this case is less pronounced as compared to the trilayer-system with 30 nm Py layer thickness described above. The corresponding hysteresis curve shows a vortex-like loop shape. Consequently, the dipolar interaction triggers the nucleation of a vortex state. However, when an identical Py-dot without Co is relaxed from the paramagnetic state to its energetic ground state also a vortex is found. Therefore it can be concluded that the dipolar coupling to the Co layer does not 'create' a vortex state, but rather that it lowers the energy barrier for relaxing to the more favorable vortex state [42].

In the third case, shown in Fig. 1 (e) and (f), a 360°-DW appears upon dipolar coupling with the Co layer. Panel (e) shows the reference hysteresis loop for the case $t_{Py} = 9$ nm without Co. The square loop-shape with the corresponding spin structure imaged at $H$ = -0.3 kOe [upper image] and H = -0.45 kOe [lower image] evidence a one step switching of the Py layer. However, in the dipolar coupled case with $t_{Co}$ = 3 nm a "C-state" is encountered [19, 20], which develops into a 360°-DW [panel (f) and inset for $H$ = 0.15 kOe and $H$ = -0.60 kOe]. This is also found for all smaller values of $t_{Py}$ from 3 to 6 nm.

These results can be summarized in a $t_{Py}$- $t_{Co}$- phase diagram as shown in Fig. 2. Four regions are marked, i.e., 'A': reversal via direct switching, 'B': stabilization of a vortex, 'C': triggering of a vortex and 'D': occurrence of a C-state and 360°-DW. In region 'A' the Py layer reverses by direct switching without a vortex state. In region 'B' a vortex occurs already in the isolated Py dot, but it becomes stabilized by dipolar interactions to the Co layer. In 'C' a triggering of a vortex state is found, i.e., the vortex only occurs in the presence of the Co layer. In region 'D' a 360°-DW develops from a C-state. In the mixed region 'C/D' we observe a reversal process via an intermediate and metastable C-state, which in some cases may develop into a vortex state.

In order to understand and to distinguish these scenarios, plots of the energy contributions *vs.* the applied field are considered. Fig. 3 (a-f) shows energy plots corresponding to the cases presented in Fig. 1 (a-f), i.e., (a) $t_{Py}$ = 30 nm, $t_{Co}$ = 0, (b) $t_{Py}$ = 30 nm, $t_{Co}$ = 3 nm, (c) $t_{Py}$ = 21 nm, $t_{Co}$ = 0, (d) $t_{Py}$ = 21 nm, $t_{Co}$ = 3 nm, (e) $t_{Py}$ = 9 nm, $t_{Co}$ = 0 and (f) $t_{Py}$ = 9 nm, $t_{Co}$ = 3 nm. In our case the formation of the vortex depends only on two energy terms, i.e., the demagnetization energy, $E_d$, and the Zeeman energy, $E_{Ze}$ of the entire system. Therefore, these two contributions and in addition for comparison the exchange energy are depicted.

For $t_{Py}$ = 30 nm without Co layer [Fig. 3 (a)] the nucleation of the vortex occurs at negative field values, $H$ = -0.3 kOe, in the descending hysteresis branch (solid symbols) [42]. Vortex nucleation is mainly induced by the minimization of the demagnetization energy (blue circles). The annihilation of the vortex at -1.05 kOe is due to the Zeeman energy (red squares) term, which becomes increasingly important with decreasing negative fields. In addition, one observes a slight increase in the exchange energy (black triangles) due to nucleation of the vortex. However, when a 3 nm thick Co layer is present the situation is different, as can be seen in Figure 3 (b). The hysteresis is biased by the Co layer and the vortex nucleates at a positive field of $H$ = 0.15 kOe, reducing the demagnetization energy but increasing slightly the slope of the Zeeman energy curve. The latter is due to a higher Zeeman energy compared to the case when the Py layer stayed in the single domain state for positive applied fields. Eventually the vortex state becomes energetically unfavorable in large negative applied fields and thus annihilation of the vortex occurs at $H$ = -0.9 kOe, which leads to a decrease of the Zeeman and an increase of the demagnetization energy. Coming from negative fields in the ascending branch, the vortex nucleates also at positive fields due to dipolar biasing of the Py layer. Here, both the demagnetization energy and the Zeeman energy are reduced. Moreover, for both field sweep directions one finds a slight reduction of the stray field energy before the system enters the vortex state. This can be described by the W-state mentioned above which occurs directly before vortex nucleation. In this case the two magnetic poles at the right and left hand side of the dot are displaced and thus lying closer together leading to a decreasing stray field [19, 20].

In Figure 3 (c) and (d) the case of triggering a vortex is displayed. Fig. 3 (c) shows the energy vs. $H$ for direct switching. Here, the magnetization reversal is completely determined by the Zeeman energy. Switching occurs at -0.3 kOe (+0.3 kOe) for the descending (ascending) field branch, respectively. Figure 3 (d) shows the case of a triggered vortex state. One observes a strong similarity to the case above [Fig. 3 (b)] and again the reduction of the stray field energy due to the W-state. One can conclude that the dipolar coupling to the Co layer generates the W-state in a thick enough Py layer (i.e. cases of stabilization and triggering). Lying energetically closer to the vortex state the core-nucleation is then easier compared to the case where the system is in the single domain (onion) state.

The system considered in Fig. 3 (e) reverses via direct switching and consequently shows analogous behavior to the one depicted in Fig. 3 (c). However, in the case with Co layer, the energy curves are strongly modified due to reversal via a 360°-DW as seen in Fig. 3 (f). A C-state nucleates at +0.15 kOe which reduces slightly the demagnetization energy and enhances slightly the Zeeman energy. It develops into a 360°-DW at $H$ = -0.6 kOe and at $H$ = -0.9 kOe the DW annihilates, which significantly reduces the demagnetization energy, the Zeeman energy and the exchange energy. In this case the antiparallel alignment of both layers is the energetically more favorable state. The minimum of the demagnetization energy for $H$ = -0.15 kOe corresponds to the case, where the C-state appears similar to a vortex state (without an out-of-plane component) minimizing the stray field contribution of the Py layer. Similar to the W-state, also the C-state lies energetically between the onion state and the vortex state. However in our case it develops into a 360°-DW. This indicates that it might be energetically more favorable than vortex nucleation. In the ascending field branch the C-state occurs at 0.15 kOe, which increases the demagnetization energy but reduces the *slope* of the Zeeman energy curve. The latter is due to a smaller Zeeman energy compared to the case when the Py layer stayed in the single domain state. The annihilation of the DW at $H$ = 1.05 kOe then leads to a decrease of the demagnetization energy and Zeeman energy.

In conclusion, we have studied dipolar coupled Co/IS/Py trilayer dots by use of micromagnetic simulations and constructed a phase diagram of the magnetic states occurring in the Py layer. The anisotropy of the Co is chosen in such a way that it stays always in a single-domain state during field cycling. Varying the thickness of the Py and the Co layer we observe four different regions, i.e., vortex stabilization, triggering of a vortex, occurrence of a C-state and 360°-DW and a mixed phase, where a C-State and in some cases vortex nucleation appear. The minimization of the demagnetization energy is responsible for the nucleation of a vortex in the Py. Obviously it is energetically more favorable to reduce the stray field energy inside the Py layer instead of forming a flux-closed antiparallel alignment between the Py and the Co layer. The dipolar interaction with the HFM Co layer leads to the occurrence of metastable W- and C-states in the Py layer. The W-states and the C-state are lying energetically closer to the vortex state and a 360°-DW, respectively. This reduces the energy barrier for the development of a a vortex state for thick Py layers and a 360°-DW for thin Py layers, respectively.

The authors would like to thank R. Hertel and J.-C. Lévy for fruitful discussions and the Deutsche Forschungsgemeinschaft through SFB 491 for financial support.

**References:**

[1]     T. Miyazaki and N. Tezuka, J. Magn. Magn. Mater. **139**, L231 (1995).

[2]     J.S. Moodera, L.R. Kinder, T.M. Wong, and R. Meservey, Phys. Rev. Lett. **74**, 3273 (1995).

[3]     S.S.P. Parkin, R.E. Fontana, and A.C. Marley J. Appl. Phys. **81**, 5521 (1997).

[4]     E.Y. Tsymbal, O.N. Mryasov, and P.R. LeClair, J. Phys.: Condens. Matter **15,** R109 (2003).

[5]     J.Z. Sun and D.C. Ralph, J. Magn. Magn. Mater. **320**, 1227 (2008).

[6]     I.L. Prejbeanu, M. Kerekes, R.C. Sousa, H. Sibuet, O. Redon, B. Dieny, and J. P. Nozières, J. Phys.: Condens. Matter **19,** 165218 (2007).

[7]     T. Shinjo, T. Okuno, R. Hassdorf, K. Shigeto, and T. Ono, Science **289**, 930 (2000).

[8]     R.P. Cowburn, D.K. Koltsov, A.O. Adeyeye, M.E. Welland, and D.M. Tricker, Phys. Rev. Lett. **83**, 1042 (1999).

[9]     A. Moser, K. Takano, D.T. Margulies, M. Albrecht, Y. Sonobe, Y. Ikeda, S. Sun, and E.E. Fullerton, J. Phys. D: Appl. Phys. **35**, R157 (2002).

[10]    T. Kimura, Y. Otani, H. Masaki, T. Ishida, R. Antos, and J. Shibata, Appl. Phys. Lett. **90**, 132501 (2007).

[11]  M.-S. Lee, A. Westphalen, A. Remhof, A. Schumann, and H. Zabel, J. Appl. Phys. **103**, 093913 (2008).

[12]  S. Bohlens, B. Krüger, A. Drews, M. Bolte, G. Meier, and D. Pfannkuche, Appl. Phys. Lett. **93**, 142508 (2008).

[13]  B. Van Waeyenberge, A. Puzic, H. Stoll. K.W. Chou. T. Tyliszczak. R. Hertel. M. Fähnle, H. Brückl, K. Rott, G. Reiss, I. Neudecker, D. Weiss, C.H. Back, and G. Schütz, Nature **444**, 431 (2006).

[14]  S.Y.H. Lua, S.S. Kushvaha, Y.H. Wu, K.L. Teo, and T. C. Chong, Appl. Phys. Lett. **93**, 122504 (2008).

[15]  K. Yamada, S. Kasai, Y. Nakatani, K. Kobayashi, H. Kohno, A. Thiaville, and T. Ono, Nature Materials **6**, 270 (2007).

[16]  W. Scholz, K. Yu. Guslienko, V. Novosad, D. Suess, T. Schrefl, R.W. Chantrell, and J. Fidler, J. Magn. Magn. Mater. **266**, 155 (2003).

[17]  Y. Otani, H. Shima, K. Guslienko, V. Novosad, and K. Fukamichi, phys. stat. sol. (a) **189**, 5421 (2002).

[18]  I.L. Prejbeanu, M. Natali, L.D. Buda, U. Ebels, A. Lebib, Y. Chen, and K. Ounadjela, J. Appl. Phys. **91**, 10 (2002).

[19]  J. Kin Ha, R. Hertel, and J. Kirschner, Phys. Rev. B **67**, 224432 (2003).

[20]  J. Kin Ha, R. Hertel, and J. Kirschner, Phys. Rev. B **67**, 064418 (2003).

[21]  M. van Kampen, I.L. Soroka, R. Bručas, B. Hjörvarsson, R. Wieser, K. D. Usadel, M. Hanson, O. Kazakova, J. Grabis, H. Zabel, C. Jozsa, and B. Koopmans, J. Phys.: Cond. Mat. **17**, L27 (2005).

[22]  A.A. Fraerman, B.A. Gribkov, S.A. Gusev, A. Yu. Klimov, V.L. Mironov, D.S. Nikitushkin, V.V. Rogov, S.N. Vdovichev, B. Hjörvarsson, and H. Zabel, J. Appl. Phys. **103**, 073916 (2008).

[23]   V.S. Pribag, I.N. Krivorotov, G.D. Fuchs, P.M. Braganca, O. Ozatay, J. C. Sankey, D.C. Ralph, and R.A. Buhrman, Nature Physics **3**, 878 (2007).

[24]   S. Kasai, Y. Nakatani, K. Kobayashi, H. Kohno, and T. Ono, Phys. Rev. Lett. **97**, 107204 (2006).

[25]   S.-K. Kim, K.-S. Lee, Y.-S. Yu, and Y.-S. Choi, Appl. Phys. Lett. **92**, 022509 (2008).

[26]   K.Y. Guslienko, K.-S. Lee, and S.-K. Kim, Phys. Rev. Lett. **100**, 027203 (2008).

[27]   Y. Liu, H. He, and Z. Zhang, Appl. Phys. Lett. **91**, 242501 (2007).

[28]   K.S. Buchanan, K. Yu. Guslienko, A. Doran, A. Scholl, S.D. Bader, and V. Novosad, Phys. Rev. B **72**, 134415 (2005).

[29]   Y. Choi, D.R. Lee, J.W. Freeland, G. Srajer, and V. Metlushko, Appl. Phys. Lett. **88**, 112502 (2006).

[30]   OOMMF/OXSII micromagnetic simulation framework by M. Donahue and D. Porter, http://math.nist.gov/oommf.

[31]   R. Skomski, J. Phys.: Condens. Matter **15**, R841 (2003).

[32]   M.H. Saevey, Jr and P.E. Tannenwald, J. Appl. Phys, **30**, 227S (1959).

[33]   K.-M. Wu, L. Horng, J.-F. Wang, J.-C. Wu, Y.-H. Wu, and C.-M. Lee, Appl. Phys. Lett. **92**, 262507 (2008).

[34]   R.D. McMichael and M. J. Donahue, IEEE Trans. Magn. **33**, 4167 (1997).

[35]   R.K. Dumas, K. Liu, C.- P. Li, I.V. Roshchin, and I.K. Schuller, Appl. Phys. Lett. **91**, 202501 (2007).

[36]   R.P. Cowburn, J. Phys. D: Appl. Phys. **33**, R1 (2000).

[37]   J. Mejía-López, D. Altbir, A. H. Romero, X. Batlle, I. V. Roshchin, C.-P. Li, and I.K. Schuller, J. Appl. Phys. **100**, 104319 (2006).

[38]   Z.-P. Li, O. Petracic. J. Eisenmenger, and I. K. Schuller, Appl. Phys. Lett. **86**, 072501 (2005).

[39]     J. Sort, G. Salazar-Alvarez, M.D. Baró, B. Dieny, A. Hoffmann, V. Novosad, and J. Nogués, Appl. Phys. Lett. **88**, 042502 (2006).

[40]     K.Yu. Guslienko, V. Novosad, Y. Otani, H. Shima, and K. Fukamichi, Phys. Rev. B **65**, 02441 (2001).

[41]     M. Schneider, H. Hoffmann, S. Otto, Th. Haug, and J. Zweck, J. Appl. Phys. **92**, 1446 (2002).

[42]     A.-V. Jausovec, G. Xiong, and R.P. Cowburn, Appl. Phys. Lett. **88**, 052501 (2006).

**Figure captions:**

Fig. 1. Hysteresis (Py minor) loops $M(H)$ for the cases: (a) $t_{Py}$ = 30 nm, $t_{Co}$ = 0, (b) $t_{Py}$ = 30 nm, $t_{Co}$ = 3 nm, (c) $t_{Py}$ = 21 nm, $t_{Co}$ = 0, (d) $t_{Py}$ = 21 nm, $t_{Co}$ = 3 nm, (e) $t_{Py}$ = 9 nm, $t_{Co}$ = 0 and (f) $t_{Py}$ = 9 nm, $t_{Co}$ = 3 nm. Insets show corresponding spin structures in panel (a) at $H$ = -0.15 [upper] and -0.3 kOe [lower image], in (b) at +0.3 and +0.15 kOe, in (c) at -0.3 and -0.45 kOe, in (d) at +0.3 and +0.15 kOe, in (e) at -0.3 and -0.45 kOe and in (f) at +0.15 and -0.6 kOe. The color code 'blue-white-red' corresponds to the x-component of the magnetization: (-1.0, 0, +1.0) $M_s$.

Fig. 2. Phase diagram of reversal modes depending on the Py and Co thickness. Letters mark different regions, i.e. 'A': reversal via direct switching, 'B': stabilization of a vortex, 'C': triggering of a vortex and 'D': occurrence of a 360°-DW. 'C/D' marks a mixed phase. The inset shows a schematic of the simulated trilayer dot with layers, Py, IS (=Insulator) and Co.

Fig. 3. Energy vs. applied field corresponding to the cases in Fig. 1 (a-f), i.e., (a) $t_{Py}$ = 30 nm, $t_{Co}$ = 0, (b) $t_{Py}$ = 30 nm, $t_{Co}$ = 3 nm, (c) $t_{Py}$ = 21 nm, $t_{Co}$ = 0, (d) $t_{Py}$ = 21 nm, $t_{Co}$ = 3 nm, (e) $t_{Py}$ = 9 nm, $t_{Co}$ = 0 and (f) $t_{Py}$ = 9 nm, $t_{Co}$ = 3 nm. Shown are the demagnetization energy (blue circles), Zeeman energy (red squares) and exchange energy (black triangles) for the descending (solid symbols) and ascending field branch (open symbols).

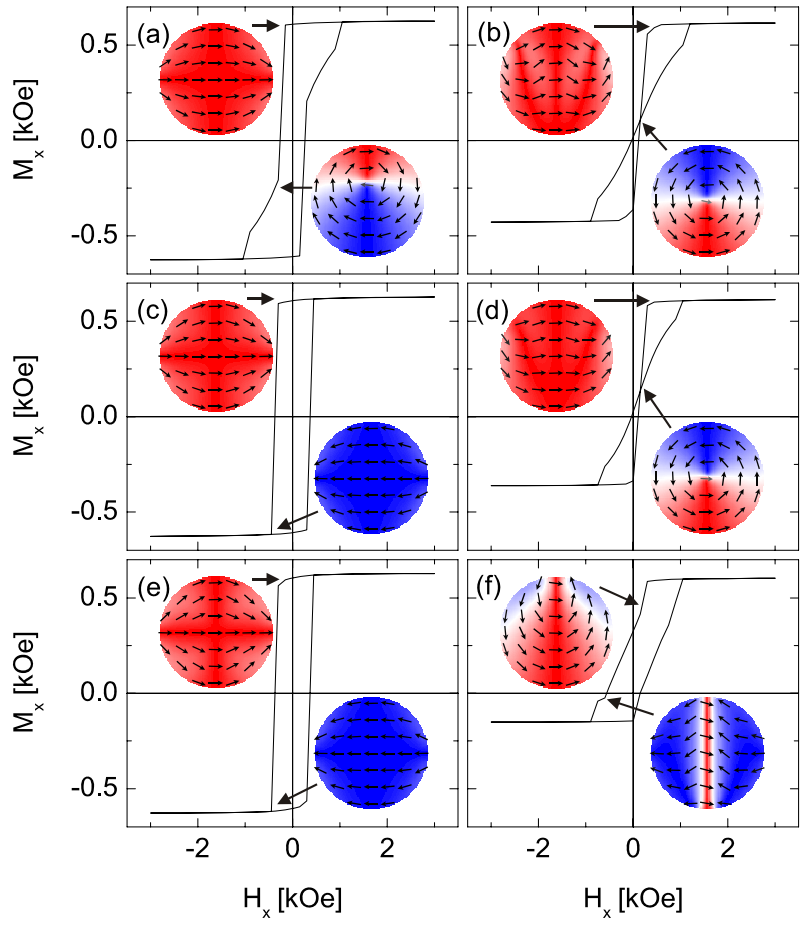

Fig. 1

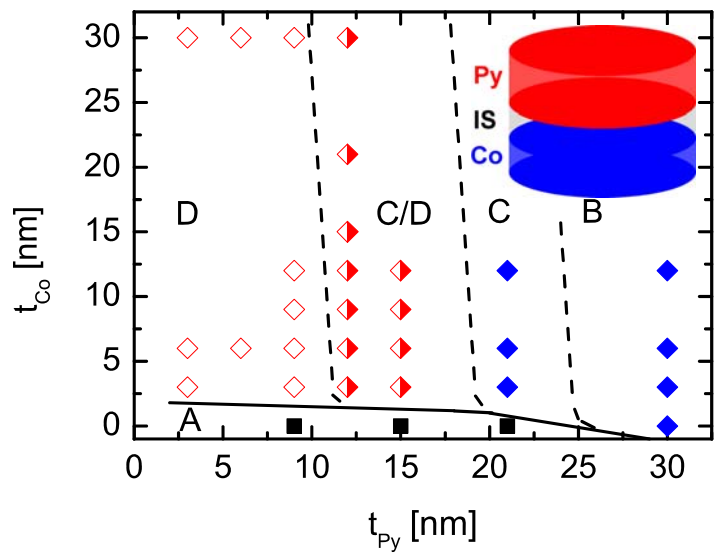

Fig. 2

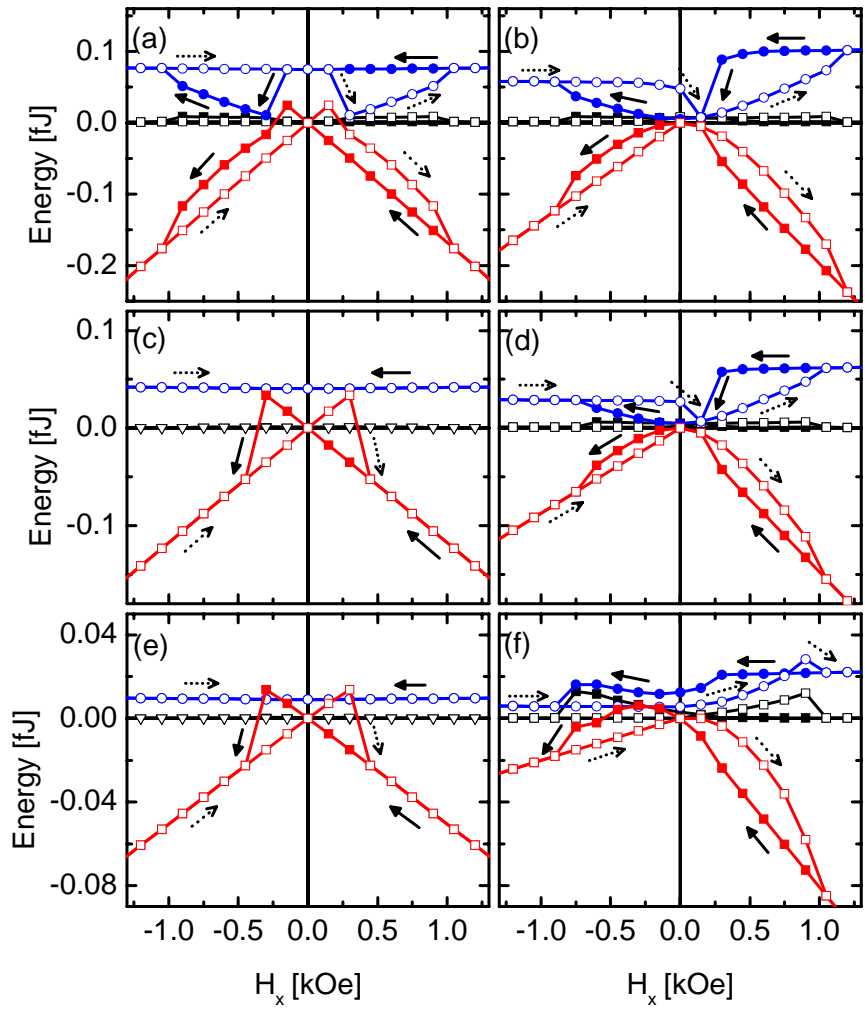

Fig. 3